\documentclass[amssymb,amsfonts,aps,prl,twocolumn,longbibliography]{revtex4-1}

\usepackage{graphicx}
\usepackage{dcolumn}
\usepackage{bm}
\usepackage[normalem]{ulem}
\usepackage{xcolor}
\usepackage{comment}
\usepackage{amsmath}
\usepackage{siunitx}
\usepackage{dsfont}
\usepackage{float}
\usepackage{placeins}
\usepackage[shortlabels]{enumitem}

\DeclareSIUnit\angstrom{\text{Å}}

\begin{document}

\title{Designer spin models in tunable two-dimensional nanographene lattices}

\author{
J. C. G. Henriques$^{1,2}$,
Mar Ferri-Cort\'es$^{3}$, 
J. Fern\'andez-Rossier$^{1,}$ 
}

\thanks{On permanent leave from Departamento de F\'isica Aplicada, Universidad de Alicante, 03690 San Vicente del Raspeig, Spain}

\affiliation{
$^1$International Iberian Nanotechnology Laboratory (INL), Av. Mestre Jos\'e Veiga, 4715-330 Braga, Portugal
}

\affiliation{
$^2$ Universidade de Santiago de Compostela, 15782 Santiago de Compostela, Spain
}

\affiliation{$^3$ Departamento de F\'isica Aplicada, Universidad de Alicante, 03690 San Vicente del Raspeig, Spain}

\date{\today}

%------------------------------------------------------------------------

%TC:ignore

\begin{abstract} 
Motivated by recent experimental breakthroughs, we propose a strategy to design two-dimensional spin lattices with  competing interactions that lead to non-trivial  emergent quantum states. We consider $S=1/2$ nanographenes with $C_3$ symmetry as building blocks, and we leverage  the potential to control both the sign and the strength of exchange with  first neighbours to build  a family of spin models. Specifically, we consider the case of a Heisenberg model in a  triangle-decorated honeycomb lattice with competing ferromagnetic and antiferromagnetic interactions whose ratio can be varied in a wide range. Based on exact diagonalization of both fermionic and spin models we predict a quantum phase transition between a  valence bond crystal of spin singlets with triplon excitations living in a Kagomé lattice and a Néel phase of effective $S=3/2$ in the limit of dominant ferromagnetic interactions.

\end{abstract}

\maketitle

%TC:endignore

%------------------------------------------------------------------------
%\section{Introduction}
Artificial lattices, created through precision engineering at the nanoscale, can serve as powerful tools for investigating emergent quantum phenomena across various physical platforms,  such as ultra-cold atoms in optical lattices\cite{gross17}, trapped ions\cite{monroe21}, Rydberg atoms\cite{scholl21}, superconducting circuits\cite{houck12},  semiconductor quantum dots\cite{hensgens17} and magnetic adatoms\cite{khajetoorians19}. These lattices mimic the periodic structures found in condensed matter systems and can help us to  emulate and understand exotic quantum states, predicted by theoretical models\cite{kitaev2006anyons,Wei2011}, that may be challenging to observe directly in natural materials
and   can play a crucial role in the development of quantum coherent nanotechnology\cite{heinrich21}.

Here, we propose to leverage the gigantic know-how of organic chemistry  using  open-shell nanographenes as building blocks to create artificial quantum matter\cite{de22}. Open shell nanographenes are planar carbon molecules with a finite spin $S$ in the ground state.  The formation of supramolecular structures, such as dimers\cite{mishra2020,krane23}, trimers\cite{du23}, rings\cite{hieulle2021}, chains\cite{Mishra2021} and two-dimensional arrays\cite{sethi2021flat, delgado23}, that preserve the open shell nature of the building blocks\cite{catarina2023}, has recently been demonstrated. This bottom-up approach permits one to design\cite{ortiz2023b,ortiz2023a,catarina2023} two-dimensional carbon crystals with very narrow bands, linear combinations  of carbon $p_z$ orbitals, and strong electron correlations, complementing the top-bottom approach of  magic angle twisted bilayer graphene \cite{cao18a,cao18b}.

In the case of one-dimensional nanographene  chains, quantum fluctuations destroy long-range order, and outstanding %stereotypical
quantum magnetism phenomena, such as the valence bond solid (VBS) \cite{Affleck1987} with  fractional edge states  in $S=1$ chains\cite{Mishra2021}, and the topological dimerized phase of the alternate bond $S=1/2$ chain\cite{chenxiao24}, have been reported.   In two dimensions,  spin lattices can display either broken symmetry phases with magnetic order and gapless spin excitations as well as  gapped quantum disordered ground states, some of which can have emergent properties, and have potential for applications in quantum information \cite{kitaev2006anyons, Wei2011}. 

The properties of both ground state and spin excitations depend crucially on the  lattice as well as the sign, symmetry and strength of the interactions therein. For instance, experiments show that  $S=3/2$ triangulenes naturally form honeycomb lattices\cite{delgado23}  and theory shows\cite{catarina2023} that these are described with non-linear Heisenberg spin models that belong to the same class of the AKLT Hamiltonian \cite{Affleck1987} with a VBS ground state. However, calculations \cite{catarina2023} show that the non-linear exchange is way too small in this system when compared with the AKLT model, which very likely will drive the system towards a Néel ground state \cite{ganesh11} with gapless magnon excitations. 

Here we pose the challenge of engineering a planar nanographene lattice that, depending on structural design parameters, can display either a broken symmetry ground state with gapless spin excitations, or quantum disordered ground state with a gapped spectrum. For that matter, we propose a system that realizes the so called star lattice \cite{Richter2004}, a triangle decorated honeycomb graph, with $S=1/2$ spins (See Figure 1). We choose as our $S=1/2$ units phenalenyl molecules \cite{reid65}, known to form open-shell dimers\cite{krane23} and trimers\cite{du23}. In each site of the honeycomb lattice we place three phenalenyl,  coupled via a  benzene linker that, as we show, induces a weak ferromagnetic exchange. Each phenalenyl in a trimer is coupled with one additional molecule from a different site of the honeycomb lattice, in a configuration known to give  antiferromagnetic exchange \cite{jacob22,krane23}.

We start our discussion with the introduction of the Hubbard model, which we shall use to describe the phenalenyl dimers and trimers considered in this work. The use of Hubbard model to model planar open-shell nanographenes is known to yield good agreement both with quantum chemistry\cite{fernandez07,ortiz19} and experimental results\cite{mishra2020,Mishra2021}.

Explicitly, the model reads\cite{Arovas2022}:
\begin{align}
    {\cal H}= \sum_{i,j,\sigma} t_{i,j} c^{\dagger}_{i\sigma}c_{j\sigma} + U\sum_i n_{i\uparrow}n_{i\downarrow},
    \label{eq:Hubbard}
\end{align}
where the indices $i,j$ run over carbon atoms (which are first and third neighbors, only), $t_{i,j}$ stands for the hopping between sites $i$ and $j$, and $U$ is the on-site Hubbard repulsion. In the rest of the paper we shall consider the first neighbor hopping to be $t=-2.7$ eV, the third neighbor hopping to be $t_3 = t/10$ and we set the Hubbard repulsion to $U=|t|$, which has been shown to agree well with \emph{ab initio} results in previous works \cite{ortiz22, catarina2023}. The operators $c^{\dagger}_{i\sigma}$ ($c_{i\sigma}$) represent the creation (annihilation) of an electron in site $i$ with spin projection along a quantization axis $\sigma=\uparrow,\downarrow$, and $n_{i\sigma} = c^{\dagger}_{i\sigma}c_{i\sigma}$ is the corresponding number operator.

For a phenalenyl molecule, due to its sublattice imbalance of 1 (one sublattice contains one additional atom when compared with the other), a single mode at zero energy is expected in its single-particle spectrum \cite{Sutherland1986}. Accounting for the Hubbard repulsion, Lieb's theorem \cite{Lieb1989} predicts a many-body ground state with spin $S=1/2$, in agreement with the Ovchinnikov rule \cite{ovchinnikov78}. This has been verified experimentally \cite{turco23} and theoretically \cite{su2019, ahmed2022phenalenyl} in previous works.

\begin{figure}
    \centering
    \includegraphics[width=\columnwidth]{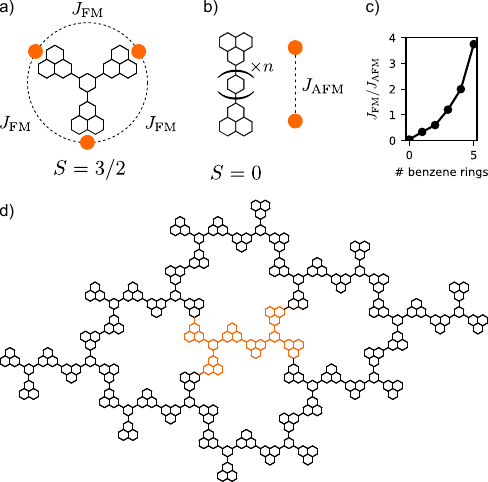}
    \caption{Schematic representation of a phenalenyl trimer a), and dimer b) with different number of benzene rings, $n$, between the two molecules; the effective $S=1/2$ spin models, and the predicted spin of the many-body ground state according to Lieb's theorem are also shown; c) Ratio of the trimer ferromagnetic and the dimer antiferromagnetic exchange for different number of benzene rings in the dimer (results obtained from a CAS calculation with the 4 least energetic orbitals in the Hilbert space); d) Depiction of a fraction of a 2D crystal obtained by combining phenalenyl trimers; the unit-cell is marked in orange.}
    \label{fig:1}
\end{figure}

When phenalenyl molecules are combined to form supra-molecular structures, such as dimers or trimers (see Fig. \ref{fig:1}a and b), Lieb's theorem\cite{Lieb1989} predicts that the many-body ground state will have a spin which is larger or smaller than $S=1/2$ depending on how the molecules are connected. In particular, if three phenalenyl molecules are linked together via a central benzene ring, preserving $C_3$ symmetry, Lieb's theorem states that the three $S=1/2$ spins couple ferromagnetically, leading to a many-body ground state with $S=3/2$; this is a consequence of the sublattice imbalance increasing to 3, and can be microscopically explained via Hund's ferromagnetic exchange between the zero modes hosted by the the three phenalenyl building blocks \cite{ortiz19, jacob22}. 

In order to estimate the magnitude of the effective ferromagnetic interaction that leads to a $S=3/2$ ground state, we solve the Hubbard model for the trimer of Fig. \ref{fig:1}a using the Configuration Interaction (CI) method in the Complete Active Space (CAS) approximation, where the Hilbert space is restricted to the three single particle states at zero energy \cite{ortiz19, supp}. Doing so, we find that, in agreement with Lieb's theorem, the molecule presents a quartet ground state corresponding to a $S=3/2$ spin; the first excited state is also a quartet, made of two degenerate doublets. Based on the low energy spectrum of the CAS calculation, we postulate a spin model to describe this system, consisting of three $S=1/2$ spins coupled ferromagnetically all-to-all, with an exchange coupling given approximately by $J_\textrm{FM} \approx -3$ meV. 

Contrarily to the trimer, the phenalenyl dimer has no sublattice imbalance, and as a consequence its many-body ground state is a singlet. Microscopically, this implies that intermolecular exchange is antiferromagnetic \cite{jacob22,krane23}, and favors low spin ground states. Performing a CAS calculation for the dimer (where only the zero modes of the two phenalenyl are accounted for in the Hilbert space), one indeed finds a singlet ground state split by approximately $63$ meV from a triplet excited state \cite{supp}, in qualitative agreement with recent experiments\cite{krane23}; at higher energies, well separated from these low energy states, other excitations appear. The results of the multiconfigurational calculation lead us to describe this molecule as two antiferromagnetically coupled spin $1/2$ with an antiferromagnetic exchange of $J_\textrm{AFM} \approx 63$ meV. Importantly, by introducing benzene rings between the molecules, our calculations show that the magnitude of the antiferromagnetic intermolecular exchange is reduced dramatically (see Fig. \ref{fig:1}c). This reduction has also been reported in the case of $S=1$ triangulene dimers\cite{mishra2020}.

\begin{figure}
    \centering
    \includegraphics[scale = 1]{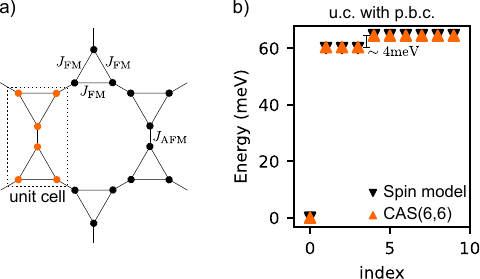}
    \caption{a) Schematic representation of the spin model for the crystal, forming a star lattice. The dashed rectangle highlights the unit cell of the crystal b) Comparison between the results obtained from CAS(6,6) and exact diagonalization of the spin model for the unit cell (u.c.) of the crystal with periodic boundary conditions (p.b.c.)}
    \label{fig:spin_model}
\end{figure}

Based on the analysis we have just performed, one sees that if two phenalenyl trimers are combined in the way depicted in Fig. \ref{fig:1}d, then it is possible to build a 2-dimensional nanographene crystal with competing ferromagnetic and antiferromagnetic interactions in the unit cell. Since we have argued that the building blocks of such a crystal are well described with spin models, it is only natural to extend the same reasoning to the crystal. Hence, we propose the following spin $1/2$ Heisenberg  Hamiltonian in the star lattice to describe the structure depicted in Fig. \ref{fig:1}d:
\begin{equation}
    H=\sum_{i,i'} J_\textrm{FM} 
    \vec{S}_i\cdot\vec{S}_{i'}
    + \sum_{i,j}J_\textrm{AFM}\vec{S}_i\cdot\vec{S}_{j} \label{eq: spin model}
\end{equation}
where $i$ and $i'$ are first neighbours belonging to the same trimer, while $i$ and $j$ are first neighbours which belong to adjacent trimers. The values of $J_\textrm{FM}$ and $J_\textrm{AFM}$ are obtained from our Hubbard-CAS calculations for the trimer and dimer, respectively.
In Fig. \ref{fig:spin_model}a we depict the spin model for the crystal, where the geometry of a star lattice is evident. A variation of this model, where Kitaev exchange is also included was considered in Ref. \cite{Peru2023}.

In order to validate the proposed spin model, we %shall now
compared the results obtained from exact diagonalization for the unit cell of the crystal with periodic boundary conditions, to the ones obtained with CAS for the same structure for the case of $n=0$ benzene spacers in the dimer. An excellent agreement between the two approaches is seen in  Fig. \ref{fig:spin_model}b, thus validating the proposed spin model. In the CAS calculation only the six single particle states closer to zero energy were considered, as these are expected to give the biggest contribution to the low energy part of the many-body spectrum. Even though we account only for exchange interactions between first neighbors, we could, in principle, consider interactions between second neighbors and beyond. For the present system, the great agreement between the spin and fermionic models (see Fig. \ref{fig:spin_model}b) shows that interactions beyond first neighbors are very small.  However, in other  star-lattice systems with second neighbor exchange that promotes frustration, for instance, new phases could appear.

The proposed star lattice model has two competing interactions, $J_{\textrm{FM}}$ and $J_{\textrm{AFM}}$, whose ratio can be controlled, in the proposed system, by changing the number of benzene spacers, $n$, in the phenalenyl dimers. We now discuss two limiting cases of the model, $J_{\textrm{FM}}=0$ and $J_{\textrm{AFM}}=0$. In the former, the star lattice becomes a Kagomé lattice of independent pairs of antiferromagnetically coupled phenalenyl molecules. The ground state wave function in this cases is a valence bond crystal (VBC), with an excited manifold of  $3N$ degenerate $S=1$ excitations with energy $J_{\textrm{AFM}}$, where $N$ is the number of unit cells  in the crystal. When $J_{\textrm{FM}}$ is turned on, but remains much smaller than $J_\textrm{AFM}$, the ground state remains close to a VBC and the $S=1$ excited states acquire a dispersion, forming triplon bands (more on this below); this explains the energy splitting seen in the excited manifold in Fig. \ref{fig:spin_model}b. Solving the model analytically, one finds the splitting of the first excited manifold to be given by $3J_\textrm{FM}/2$, in good agreement with the numerical result. In the  opposite limit, i.e. $J_{\textrm{AFM}}=0$, we find $2N$ independent trimers with $S=3/2$ spin. When $J_{\textrm{AFM}}$ is turned on, we are left with a honeycomb lattice of $S=3/2$ spins with antiferromagnetic interactions. This broken symmetry  limit has been discussed by Owerre \cite{owerre17}, who found topological magnons.

The star lattice Hamiltonian has been studied before in the case where both interactions are antiferromagnetic \cite{Richter2004, yang10, jahromi18,Reingruber2023}, where  spin frustration in the triangles can promote spin-liquid behaviour, consistent with recent experimental observations \cite{sorolla2020synthesis}. The case with intratriangle antiferromagnetic and intertriangle ferromagnetic exchange couplings has also been studied  \cite{Ran2018}.
In our case, however, the interactions within each triangle are ferromagnetic instead, and antiferromagnetic exchange links different triangles. To infer the properties of this system we now exactly diagonalize the spin model for a ring made of six trimers, containing a total of 18 spins, with periodic boundary conditions (see Fig. \ref{fig:spin_model}a). We focus on two aspects, the nature of the ground state wave function and the low energy excitation spectrum.

\begin{figure}
    \centering
    \includegraphics[scale = 1]{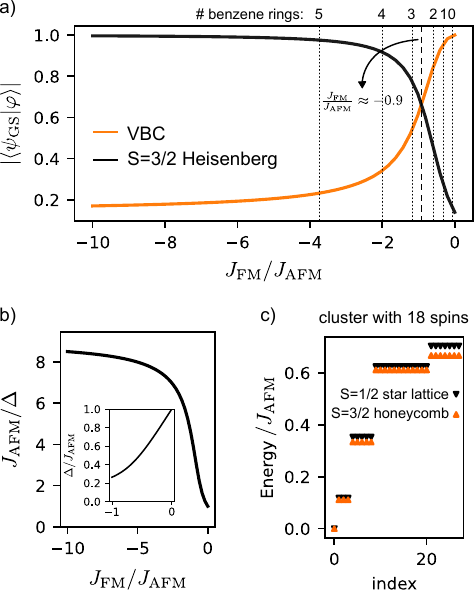}
    \caption{
    a) Overlap between the wave function of the ground state of a ring of 6 trimers, with a total of 18 spins, with p.b.c., $| \psi_\textrm{GS} \rangle$, and the wave functions of the valence bond crystal (VBC) and the $S=3/2$ Heisenberg model ground state for the same geometry. The dashed vertical line marks the point where the two overlaps are identical, and the dotted lines refer to the values of $J_\textrm{FM}/J_\textrm{AFM}$ for different number number of benzene spacers in the phenalenyl dimers (c.f. Fig. \ref{fig:1}c); 
    b) Energy of the first excitation, $\Delta$, as a function of $J_\textrm{FM} / J_\textrm{AFM}$; 
    c) Exact diagonalization of spin model the 6 trimer ring with $J_\textrm{FM} = -10 J_\textrm{AFM}$ for the $S=1/2$ star lattice and $\mathcal{J}^{\textrm{eff.}}_{S = 3/2} = J_\textrm{AFM} / 9$ for the $S=3/2$ honeycomb geometry.
    }
    \label{fig:cluster}
\end{figure}

To gain insight about the nature of the ground state of this system, we compute the overlap between its wave function and another two particularly relevant states: i) the wave function of the VBC; ii) the ground state of the $S=3/2$ antiferromagnetic Heisenberg model. For the first case, we simply build the VBC wave function by solving the spin model with $J_\textrm{FM} = 0$, and then compute the overlap with the solution for several values of $J_\textrm{FM}/J_\textrm{AFM}$. Regarding the second case, we solve the $S=3/2$ Heisenberg model which we then project into a spin-1/2 basis, thus allowing the computation of the overlap with the wave function of our system;  the results are displayed in Fig. \ref{fig:cluster}a. 
There, we see that, as expected, the overlap between the exact ground state wave function and the one from the VBC starts at 1 for $J_\textrm{FM} = 0$, and quickly decreases as the ratio of ferro- and antiferromagnetic interactions increases. At odds, with this, the overlap with the $S=3/2$ Heisenberg model wave function starts at approximately 0.14  and increases as $|J_\textrm{FM}/J_\textrm{AFM}|$ grows. In the limit $|J_\textrm{FM}| \gg J_\textrm{AFM}$ the overlap approaches 1, as expected given the physical picture presented before. 
Crucially, by controlling the molecular linking groups it seems possible to induce a quantum phase transition in the system, as its ground states changes from a quantum disordered one in the VBC limit, to a broken symmetry one in the Heisenberg limit (the magnetization map of the mean-field Hubbard ground state is shown in \cite{supp}).

In Fig. \ref{fig:cluster}b and c we focus on the excitation spectrum obtained via exact diagonalization of the spin model for the 18 spin cluster we have just considered. For all considered values of $J_\textrm{FM}/J_\textrm{AFM}$ we find a singlet ground state followed by a triplet excitation. In panel b we plot the energy of the first excited manifold, $\Delta$, as a function of the ratio of ferro- and antiferromagnetic interactions. In the limit $|J_\textrm{FM}|/J_\textrm{AFM} \ll 1$, where the system is close to a VBC with a quantum disordered ground state, the gap of the system is essentially $J_\textrm{AFM}$ and the first set of excited states corresponds to a set of almost degenerate triplets, slightly split due to the finite $J_\textrm{FM}$. As $|J_\textrm{FM}|/J_\textrm{AFM}$ increases, and the system transitions from a VBC to a $S = 3/2$ Heisenberg model, the gap of the first excitation decreases sharply, and saturates at ${\cal J}_{S = 3/2}^{\textrm{eff.}} = J_\textrm{AFM}/9$.  The finite value of this gap is due to  the finite size of our simulation lattice. The relation between ${\cal J}_{S = 3/2}^{\textrm{eff.}}$ and $J_{\textrm{AFM}}$ is identical to the one expected from a classical model \cite{supp}. In Figure \ref{fig:cluster}c we compare the low energy spectrum of the $S=1/2$ star lattice, using $J_{\textrm{FM}}=- 10  J_{\textrm{AFM}}$, with the energy spectrum of a $S=3/2$ honeycomb Heisenberg model with exchange ${\cal J}_{S = 3/2}^{\textrm{eff.}}$. There one finds that the two data set are in good agreement; we have verified that as we approach the limit $|J_\textrm{FM}|/J_\textrm{AFM} \rightarrow \infty$ an exact agreement appears \cite{supp}.

\begin{figure}
    \centering
    \includegraphics[width = \linewidth]{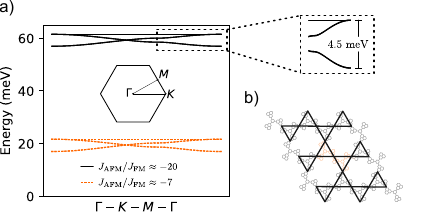}
    \caption{(a) Triplon dispersion obtained with the mean-field self-consistent bond operator formalism. Solid black lines show the bands for the case with $J_\textrm{AFM} = 63$ meV and dashed orange lines refer to the case with $J_\textrm{AFM} = 20$ meV; in both cases $J_\textrm{FM}  = -3$ meV; (b) Schematic representation of the effective Kagomé lattice formed by the phenalenyl pairs which link different trimers.}
    \label{fig:Fig4}
\end{figure}

We now focus on the  excitations in the small $|J_\textrm{FM}|/J_\textrm{AFM}$ limit. These excitations are triplons, which correspond to $S=1$ excitations that propagate in the Kagomé lattice defined by the bonds of the honeycomb lattice of  antiferromangetically coupled phenalenyls. In this limit, we can employ the mean-field self-consistent bond-operator formalism introduced by Sachdev and Bhatt \cite{sachdev1990bond, gopalan1994spin, normand1996electronic}. Within this framework, we assign a singlet and three triplet creation operators to each antiferromagnetically coupled dimer of the unit cell. Expressing the Hamiltonian of the spin model in terms of these bosonic bond operators, and diagonalizing it via a paraunitary transformation \cite{pires2021theoretical}, we find the triplon dispersion depicted in Fig. \ref{fig:Fig4} \cite{supp}. These bands follow the usual dispersion of Kagomé systems, with two dispersive bands, and a single flat one. This set of three bands is centered approximately at the energy of the antiferromagnetic exchange, and the bandwidth is controlled by the ferromagnetic coupling. Reducing the strength of the antiferromagnetic coupling, while still keeping it larger than the ferromagnetic interaction, simply leads to a downwards shift of the band structure; further decreasing $J_\textrm{AFM}$ would originate an instability in the triplon spectrum, indicating the transition from a quantum disordered ground state to a broken symmetry one.

In summary, we  propose a strategy to create artificial  two-dimensional spin lattices with tunable competing interactions using nanographenes. Specifically, we have shown that phenalenyls can be used as building blocks to realize the $S=1/2$ star lattice model with competing ferro- ($J_\textrm{FM}$) and antiferromagnetic ($J_\textrm{AFM}$) couplings. We have shown that the figure of merit of this model, the ratio  $J_\textrm{AFM}/J_\textrm{FM}$, can be tuned in a range that permits one to cross the transition between two very different phases, a $S=1/2$ valence bond crystal state with gapped triplons,  and a $S=3/2$ Néel ordered phase. Experimentally, finite size nanographene lattices are to be expected, and the VBC phase should have in gap edge state \cite{delgado23}, analogous to those of the Haldane chains. Other materials \citep{sorolla2020synthesis, zheng2007star} where the star lattice can be realized have previously been reported, but these lack the tunability offered by nanographenes.´ Our results illustrate the potential of nanographene based artificial lattices to  apply  the gigantic power of synthetic organic chemistry  to engineer non-trivial quantum spin states.

%\section*{Acknowledgements}
%TC:ignore
{\em Acknowledgements. }
We acknowledge Ant\'onio Costa and Bruno Murta for
fruitful discussions and  support with numerical calculations. %
MFC
acknowledges funding from Generalitat Valenciana
(CIACIF/2021/434).
J.F.-R., J.C.G.H. %and A.C 
acknowledge financial support from 
%1
FCT (Grant No. PTDC/FIS-MAC/2045/2021),
%2
SNF Sinergia (Grant Pimag,  CRSII5-205987) and 
%3
the European Union (Grant FUNLAYERS- 101079184). J.F.-R. acknowledges funding from
%
%FEDER/Junta de Andaluc\'ia, %--- Consejer\'ia de Transformaci\'on Econ\'omica, Industria, Conocimiento y Universidades,
%(Grant No. P18-FR-4834), 
% 4
Generalitat Valenciana (Prometeo2021/017
and MFA/2022/045)
%5
and
MICIN-Spain (Grants No. PID2019-109539GB-C41 and PRTR-C1y.I1) 

%TC:endignore

%\appendix 

%\bibliographystyle{apsrev4-1}
%\bibliography{bibshort}


\end{document}